\documentclass[11pt]{article}
\pdfoutput=1
\usepackage{dcolumn}% Align table columns on decimal point
\usepackage{bm}% bold math

\usepackage{graphicx}% Include figure files
\usepackage{amssymb,amsmath}
\usepackage{multirow}
\usepackage{cite,color,url}
\usepackage[colorlinks=true
,urlcolor=blue
,anchorcolor=blue
,citecolor=blue
,filecolor=blue
,linkcolor=blue
,menucolor=blue
,linktocpage=true
,pdfproducer=medialab
,pdfa=true
]{hyperref}

\usepackage{epsfig,psfrag,rotating,soul}
\usepackage{rotfloat}

\evensidemargin \oddsidemargin
\allowdisplaybreaks

\psfrag{lk}{$l_k$}
\psfrag{lm}{$l_m$}
\psfrag{blk}{$\bar {l}_k$}
\psfrag{blm}{$\bar {l}_m$}

\def\thefootnote{\fnsymbol{footnote}}

\newcommand{\mb}{\ensuremath{m_b}}
\newcommand{\mt}{\ensuremath{m_t}}
\newcommand{\rsm}{\ensuremath{R^{\rm SM}}}
\newcommand{\rmssm}{\ensuremath{R^{\rm MSSM}}}

\newcommand{\mtau}{\ensuremath{m_\tau}}

\newcommand{\ta}{\ensuremath{\tan\alpha}}

\newcommand{\ma}{\ensuremath{M_{A}}}
\newcommand{\mz}{\ensuremath{M_{Z}}}
\newcommand{\tb}[1][]{\ensuremath{\tan^{#1}\!\beta}}

\newcommand{\Dmb}[1][]{\ensuremath{\Delta\mb^{#1}}}
\newcommand{\Dmtau}{\ensuremath{\Delta m_\tau}}
\newcommand{\Dmf}[1][]{\ensuremath{\Delta m_f^{#1}}}
\newcommand{\TeV}{\ensuremath{\,{\rm TeV}}}
\newcommand{\GeV}{\ensuremath{\,{\rm GeV}}}

\let\OLDthebibliography\thebibliography
\renewcommand\thebibliography[1]{
  \OLDthebibliography{#1}
  \setlength{\parskip}{0pt}
  \setlength{\itemsep}{0pt plus 0.3ex}
}

\begin{document}
\thispagestyle{empty}

\vspace*{2.5cm}

\vspace{0.5cm}

\begin{center}

\begin{Large}
\textbf{\textsc{Discriminating between SUSY and Non-SUSY Higgs Sectors through the Ratio $H \to b \bar b / H \to \tau^+ \tau^-$ with a 125 GeV Higgs boson}}
\end{Large}

\vspace{1cm}

{\sc
E. Arganda$^{1}$%
\footnote{\tt \href{mailto:ernesto.arganda@unizar.es}{ernesto.arganda@unizar.es}}%
, J.~Guasch$^{2, 3}$%
\footnote{\tt \href{mailto:jaume.guasch@ub.edu}{jaume.guasch@ub.edu}}%
, W.~Hollik$^{4}$%
\footnote{\tt \href{mailto:hollik@mpp.mpg.de}{hollik@mpp.mpg.de}}%
, S.~Pe\~naranda$^{1}$%
\footnote{\tt \href{mailto:siannah@unizar.es}{siannah@unizar.es}}%
}

\vspace*{.7cm}

{\sl
$^1$Departamento de F\'{\i}sica Te\'orica, Facultad de Ciencias,\\
Universidad de Zaragoza, E-50009 Zaragoza, Spain

\vspace*{0.1cm}

$^2$Departament de F\'{\i}sica Fonamental,\\
Universitat de Barcelona, Diagonal 645, E-08028 Barcelona, Catalonia, Spain

\vspace*{0.1cm}

$^3$Institut de Ci\`encies del Cosmos (ICC),\\
Universitat de Barcelona, Diagonal 645, E-08028 Barcelona, Catalonia, Spain

\vspace*{0.1cm}

$^4$Max-Planck-Institut f\"ur Physik,\\
F\"ohringer Ring 6, D-80805 M\"unchen, Germany

}

\end{center}

\vspace*{0.1cm}

\begin{abstract}
\noindent
It is still an open question whether the new scalar particle discovered at the LHC with a mass of 125 GeV is the SM Higgs boson or it belongs to models of new physics with an extended Higgs sector, as the MSSM or 2HDM. The ratio of branching fractions $R$ = BR($H \to b \bar b$)/BR($H \to \tau^+ \tau^-$) of Higgs boson decays is a powerful tool in order to distinguish the MSSM Higgs sector from the SM or non-supersymmetric 2HDM. This ratio receives large renormalization-scheme independent radiative corrections in supersymmetric models at large $\tan\beta$, which are insensitive to the SUSY mass scale and absent in the SM or 2HDM. Making use of the current LHC data and the upcoming new results on Higgs couplings to be reported by ATLAS and CMS collaborations and in a future linear collider, we develop a detailed and updated study of this ratio $R$ which improves previous analyses and sets the level of accuracy needed to discriminate between models.
\end{abstract}

\def\thefootnote{\arabic{footnote}}
\setcounter{page}{0}
\setcounter{footnote}{0}

\newpage

\section{Introduction}
\label{intro}

At present, it is still an open question in the high energy physics community 
whether the discovered new scalar particle at the Large Hadron Collider (LHC)~\cite{HiggsDiscovery} 
is actually the Higgs boson of 
the Standard Model of particle physics (SM). This new particle seems
to behave as the SM Higgs boson, and
the most recent combined measurement of its mass by the ATLAS and CMS 
collaborations set $m_{H^\text{SM}} =$ 125.09 $\pm$ 0.21 (stat.) $\pm$ 0.11 (syst.) GeV~\cite{Aad:2015zhl}. 
However, many more measurements and data will be needed to extract 
reliable conclusions. 
It is worth noticing that the study of perturbativity and stability of
the SM Higgs boson potential suggests that, given the measured Higgs
boson mass, new physics must be present before the Planck scale~\cite{Planck}.
Apart from the introduction of new particles, extensions of the SM
scalar sector may affect the properties
of the SM-like Higgs boson discovered at the LHC. 
Experimental data is being used to constrain these extensions.
Among the minimal extensions of the SM is the inclusion of additional
Higgs bosons. 
In the two-Higgs-doublet models (2HDM) one additional Higgs doublet is introduced
and five physical Higgs bosons are obtained~\cite{Hunter}: two CP-even scalars ($h$ and $H$),
one CP-odd scalar ($A$), and a charged Higgs pair ($H^{\pm}$); 
being the lightest Higgs boson very similar to the SM one in the
so-called {\it{decoupling limit}}~\cite{Gunion:2002zf}. 
The minimal supersymmetric standard model (MSSM)~\cite{MSSM}, 
one of the most-predictive frameworks 
beyond the SM, also contains two Higgs doublets with a light neutral scalar boson
compatible with the existing measurements, including the recently discovered Higgs boson.
In this letter we approach the question of the existence of an extended Higgs structure beyond 
the SM by investigating the neutral Higgs sector of various types of
models. 

We consider in this work the ratio of
branching ratios of a neutral Higgs boson $H$~\cite{Guasch:2001wv},
\begin{equation}
  \label{eq:Rdef}
  R=\frac{\text{BR}(H \to b \bar b)}{\text{BR}(H \to \tau^+ \tau^-)}\,\,,
\end{equation}
analyzing in detail the Yukawa-coupling effects
and their phenomenological consequences. At leading order, in either 
the SM, the 2HDM or the MSSM, this ratio
is given by just the ratio of squared  (running) masses:
\begin{equation}
  \label{eq:RteoSM}
  R=3\frac{m_b^2(Q)}{m_\tau^2(Q)}\ \ .
\end{equation}
However, this ratio receives large renormalization-scheme independent radiative
corrections in supersymmetric (SUSY) models at large $\tan \beta$,
the ratio of the vacuum expectation values. These corrections are
insensitive to the SUSY mass scale ($M_\text{SUSY}$) and absent in the SM or 2HDM. Therefore,
this ratio is a discriminant quantity between SUSY and 
non-SUSY models. The leading radiative
corrections to this ratio can be cast into an effective Yukawa SUSY
coupling $h_f$, and summarized in a simple correction factor 
$\Dmf$~\cite{Guasch:2001wv,Hall:1993gn,Carena:1994bv},
thus for a down-type quark or a charged lepton one can write:
\begin{equation}
  \label{eq:deffhb}
  h_f=\frac{m_f(Q)}{v_1} \frac{1}{1+\Dmf}=
      \frac{m_f(Q)}{v \cos\beta}\frac{1}{1+\Dmf}, \quad \quad
    v= (v_1^2+v_2^2)^{1/2} \,.
\end{equation}
Here $m_f(Q)$ is the running fermion mass, $v_1$ and $v_2$ are the
vacuum expectation values ({\em{v.e.v.s}}) of the two Higgs doublets; being $v_1$ the one
giving mass to down-type quarks and charged leptons,
$\tb=\frac{v_1}{v_2}$ is the ratio of the {\em{v.e.v.s}} and $v=
(v_1^2+v_2^2)^{1/2}$ is the SM {\em{v.e.v}}.
This expression includes all possible $\tb$ enhanced corrections of the type
$(\alpha_{(s)}\tb)^n$~\cite{Carena:1999py} 
correctly resumed. The leading part of the (potentially) non-decoupling
contributions proportional to the trilinear soft-SUSY-breaking
coupling $A_f$ can be absorbed in the definition of
the effective Yukawa coupling at low energies and only subleading
effects survive~\cite{Guasch:2003cv}. Therefore,
expression~(\ref{eq:deffhb}) contains all leading potentially large
radiative effects.
The resummation of the two-loop dominant corrections for large values of
$\tan\beta$ has been calculated in~\cite{2loop-resummation}. 

The interplay between Higgs physics and SUSY, with the inclusion of radiative corrections, 
has been extensively discussed in the literature, see e.g.~\cite{Guasch:2001wv,Hall:1993gn,Carena:1994bv,Carena:1999py,Guasch:2003cv,2loop-resummation,Dabelstein:1995js,Coarasa:1996qa,Coarasa:1995yg,Carena:1999bh,Carena:1998gk,Babu:1998er,Pierce:1996zz,Haber:2000kq,CDHPTdeTr,Heinemeyer:2001pq,higgsradcor,Heinemeyer:2000fa}. 
It is also well known that the SUSY radiative corrections 
to the couplings of the Higgs bosons to bottom quarks can be 
significant for large values of $\tan \beta$, and that they do not decouple in
the limit of a heavy supersymmetric 
spectrum~\cite{Coarasa:1996qa,Coarasa:1995yg,Guasch:2003cv ,Carena:1999bh,Carena:1998gk,Babu:1998er,Carena:1999py,Pierce:1996zz,
Carena:1994bv,Hall:1993gn,Haber:2000kq,CDHPTdeTr},
opposite to their behavior in electroweak gauge boson physics~\cite{DHP}.
The partial decay width $\Gamma(h\to b \bar{b})$ 
of the lightest supersymmetric neutral Higgs particle has received particular attention. The complete
one-loop corrections have been studied in~\cite{Dabelstein:1995js}, and 
comprehensive studies of the one- and two-loop SUSY-QCD corrections are also 
available in~\cite{Coarasa:1995yg} and~\cite{Heinemeyer:2000fa}. 
The effective Lagrangian description of the $hb\bar{b}$ vertex and the
implications for Higgs-boson searches from SUSY effects
can be found in~\cite{Hall:1993gn,Guasch:2003cv,Carena:1999py,Carena:1998gk,Carena:1999bh}.
The decoupling properties
of the SUSY-QCD corrections to $\Gamma(h\to b \bar{b})$ have been
extensively discussed in~\cite{Haber:2000kq}.
On the other side, the analysis of $\text{BR}(H \to \tau^+ \tau^-)$ was
presented in~\cite{Heinemeyer:2001pq}. The observable $R$, as the ratio
of the two last mentioned processes, has been also analyzed in~\cite{Guasch:2001wv} and~\cite{Babu:1998er}.
Some recent analyses of these two branching ratios and other Higgs decay modes, confronting LHC data with the MSSM predictions, can be found, for example, in~\cite{MSSM-Higgs_probes}.

The ratio~(\ref{eq:Rdef}) is very interesting from both 
the experimental and the theoretical sides.
It is a clean observable, measurable in a
counting experiment, with only small 
systematic errors since most of them cancel in the ratio. The only
surviving systematic effect results from the efficiency of $\tau$- and $b$-tagging. 
From the theoretical point of view, it is independent of the production
mechanism of the decaying neutral Higgs boson and of its total 
width. Therefore, new-physics effects affecting the production cross-section do
not appear in the ratio and also this observable is insensitive to unknown
higher order QCD corrections to Higgs boson production. 
Besides, since this ratio only depends on the ratio of 
the masses~(\ref{eq:RteoSM}), there is no other parameter 
(e.g.\  \tb) that could absorb the large quantum corrections. 

As shown in~\cite{Guasch:2001wv}, 
the ratio of the Higgs boson decay rates into
$b$ quarks and $\tau$~leptons~(\ref{eq:Rdef}) 
normalized to  the Standard Model expectation $\rsm$ 
is a very efficient quantity to 
distinguish a general 2HDM from the MSSM, whose Higgs sector could 
be fully covered at the LHC~\cite{hMSSM}.
This normalized value 
is a function depending only on $\tb$, $\ta$, $\Dmb$, and
$\Dmtau$, and encoding all the genuine SUSY corrections. 
The explicit form of $\Dmb$ and $\Dmtau$ at the one-loop level can be 
obtained approximately 
by computing the supersymmetric loop diagrams at zero external momentum
($M_\text{SUSY} \gg m_b\,,m_\tau$)~\cite{Guasch:2001wv}. 
These two quantities are independent of the SUSY
mass scale $M_\text{SUSY}$ since they only depend on $\tb$ and 
the ratio $A_t/M_\text{SUSY}$~\cite{Guasch:2001wv,Carena:1999py,Carena:1998gk}.
Therefore, the conclusions about the sensitivity to the 
SUSY nature of the Higgs
sector through the analysis of the ratio $R$ are independent of the 
scale of the SUSY masses.

Nowadays, the experiments at the LHC become increasingly
sensitive  to the Higgs boson couplings. 
CMS and ATLAS have indeed performed a generic fit to Higgs-boson
coupling ratios. In order to carry out this analysis, they define 
a set of Higgs boson couplings normalized to the SM
ones, $\kappa_x\equiv g_x/g_x^{SM}$, and the production rates 
measurements give a
measurement of the coupling ratios for two particles:
$$
\lambda_{xy}\equiv \frac{\kappa_x}{\kappa_y}\equiv\frac{g_x/g_x^{SM}}{g_y/g_y^{SM}} \ \ .
$$
In the present work, we are interested in the bottom-quark and
$\tau$-lepton measurements, for which CMS and ATLAS collaborations 
provide~\cite{Khachatryan:2014jba}
\begin{equation}
\label{eq:lambdasexps}
\lambda_{bZ}^\text{CMS}=0.59_{-0.23}^{+0.22} \ \ , \ \ 
\lambda_{\tau Z}^\text{CMS}=0.79_{-0.17}^{+0.19} \ \ , \ \ 
\lambda_{bZ}^\text{ATLAS}=0.60 \pm 0.27\ \ , \ \ 
\lambda_{\tau  Z}^\text{ATLAS}=0.99_{-0.19}^{+0.23}\ \ .
\end{equation}
Besides, the expected accuracy for the measurement of the 
fundamental Higgs couplings $H b \bar b$ and $H \tau^+ \tau^-$ 
in future course of the LHC run corresponds to an uncertainty of
$10$-$13\%$ ($b$ quarks) and $6$-$8\%$ ($\tau$ leptons), 
going down to $4$-$7\%$ and $2$-$5\%$ for the high luminosity LHC (HL-LHC).
At the Linear Collider (LC) the expected uncertainty is smaller, $0.6\%$ for 
$H b \bar b$ coupling and $1.3\%$ for $H \tau^+ \tau^-$ coupling~\cite{accuracy-refs}.
In this paper we consider the present experimental results on the
Higgs boson mass and couplings in the analysis of the 
ratio~(\ref{eq:Rdef}) as well as the expected future precision,
and discuss the possibility to discriminate between models
at  various levels of future accuracy.

In section~\ref{MAtanb} we present the relevant expressions for our study
and analyze the ratio $R$ (eq.~(\ref{eq:Rdef})) in view of the present LHC data 
on Higgs boson coupling ratios as given in~(\ref{eq:lambdasexps}).
Section~\ref{sensitivities}
is devoted to the analysis of the future sensitivities of this ratio at
present and future colliders,
and the study of the potential discrimination between SUSY or non-SUSY models.
Finally, the conclusions of this work are summarized in section~\ref{conclusions}. 

\section{Analysis of present data}
\label{MAtanb}

In this section we concentrate on the analysis of the ratio $R$ defined in~(\ref{eq:Rdef})
for the cases of the lightest CP-even Higgs boson, $h$. 
For the sake of the discussion and the analysis it will be useful to
introduce the ratio $R$~(\ref{eq:Rdef}) normalized to the SM value for
equal values of the Higgs boson mass. For
a Higgs boson $\phi$ we define:
\begin{equation}
  \label{eq:Xnorm}
  X(\phi)=\frac{R(\phi)}{R^\text{SM}(m_{H^\text{SM}}=m_\phi)} \ \ .
\end{equation}
We can write this normalized ratio for each neutral MSSM Higgs boson 
in terms of
the non-decoupling quantities $\Dmb$ and $\Dmtau$ as~\cite{Guasch:2001wv}
\begin{eqnarray}
X(h)&=& \frac{\rmssm(h)}{\rsm} = \frac{(1 + \Dmtau)^2\,(-\cot\alpha \,\Dmb + \tan \beta)^2}
{(1 + \Dmb)^2\,(-\cot\alpha \,\Dmtau + \tan \beta)^2}\,, \label{eq:Rh0} \\
X(H)&=&\frac{\rmssm(H)}{\rsm} = \frac{(1 + \Dmtau)^2\,(\tan\alpha \,\Dmb + \tan \beta)^2}
{(1 + \Dmb)^2\,(\tan\alpha \,\Dmtau + \tan \beta)^2}\,, \label{eq:RH} \\
X(A)&=&\frac{\rmssm(A)}{\rsm} = \frac{(1 + \Dmtau)^2\,(\tan^2\beta - \Dmb)^2}
{(1 + \Dmb)^2\,(\tan^2\beta - \Dmtau)^2}\,. \label{eq:RA}
\end{eqnarray}
In~\cite{Guasch:2001wv}, by assuming a $\pm 21\%$
measurement of this ratio for the lightest Higgs boson 
at the LHC~\cite{ZeppenfeldGianotti}, 
it was found that one can be sensitive to the SUSY nature of the lightest Higgs
boson $h$ for \ma\  up to $\sim 1.8\TeV$ in the most favorable scenario,
being up to  $\ma\sim 500\GeV$ in some other regions. 
Nowadays, the combination of the LHC coupling
measurements of eq.~(\ref{eq:lambdasexps}) provides an experimental
determination of the normalized ratio~(\ref{eq:Xnorm})
\begin{equation}
  \label{eq:Xexp}
  X^\text{exp}=\frac{R^\text{exp}}{R^\text{SM}}
=\frac{\lambda_{bZ}^2}{\lambda_{\tau Z}^2} \,.
\end{equation}
Using the values in eq.~(\ref{eq:lambdasexps}) we obtain:
\begin{equation}
  \label{eq:XexpATLASCMS}
X^\text{CMS} = 0.56_{-0.52}^{+0.48} \ \ , \ \ 
X^\text{ATLAS} = 0.37_{-0.37}^{+0.36} \,.
\end{equation}
In this work we consider this  experimental determination and we discuss
their phenomenological consequences through the analysis of the
normalized ratio $X$~(\ref{eq:Xnorm}) in different SUSY scenarios. 
Besides, we also include in our numerical analysis a combined analysis
of CMS and ATLAS results. From the generic fit to Higgs
coupling ratios given above one can determine the values of
these coupling ratios to be
\begin{equation}
\label{eq:lambdascombined}
\lambda^\text{Combined}_{bZ} = 0.594_{-0.174}^{+0.171} \ \ , \ \
\lambda^\text{Combined}_{\tau Z} = 0.887_{-0.126}^{+0.140} \,. 
\end{equation}
We obtain these values by
using the procedure for combination of results described
in~\cite{Barlow:2004wg}.
As a consequence we get a value for the ratio of 
\begin{equation}
  \label{eq:Xcombined}
X^\text{Combined}=\frac{R^\text{Combined}}{R^\text{SM}} = 0.45_{-0.30}^{+0.29} \, .  
\end{equation}
The one-standard deviation (68\% C.L.) favored bands on
$X$~(\ref{eq:Xnorm}) are:
\begin{equation}
  \label{eq:Xlimits}
 0.04< X^\text{CMS}<1.04 \ \ , \ \ 0 < X^\text{ATLAS}<0.73 \ \ , \ \
 0.15< X^\text{Combined}<0.74 \,.
\end{equation}
While the CMS result includes the SM value ($X=1$) in its favored
region, the ATLAS and our combined results disfavor the SM (at 68\% C.L.).
SUSY can provide the necessary corrections to bring the predicted
theoretical value of $X$ inside the ATLAS favored band.

For the theoretical numerical analysis, we consider different SUSY scenarios, 
by checking that those scenarios are compatible with the 
present experimental value of the Higgs boson mass, 
$m_{H^\text{SM}} = 125.09 \pm 0.21 \text{ (stat.)} \pm 0.11 \text{ (syst.)} \text{ GeV}$~\cite{Aad:2015zhl}.
The Higgs boson mass is computed by using 
{\tt FeynHiggs 2.11}~\cite{Heinemeyer:1998yj}. 
For completeness, first we consider the four scenarios analyzed 
in~\cite{Guasch:2001wv} and we find that the SUSY spectra defined 
in these scenarios provide a Higgs boson mass value not compatible 
with the present experimental result. The only exception is the 
scenario with $\mu<0$ $A_t>0$, in which we obtain
that $m_h$ is around $122 \GeV$. For the purpose of making contact 
with the previous results, 
we include this scenario in the following discussion. 
Besides, we choose  SUSY spectra as defined in~\cite{Carena:2013qia} for 
the $m_h^\text{mod+}$, $m_h^\text{mod-}$, light-stop, and light-stau 
scenarios in the MSSM, which are compatible
with the Higgs boson mass of the observed signal at the LHC, and the 
benchmark scenario 2392587 of the phenomenological MSSM (pMSSM)~\cite{Cahill-Rowley:2013gca},
a general version of the R-parity conserving MSSM with 20 input parameters. The SM
parameters are fixed to be: $\mt=173.21\GeV$, $\mb=4.18\GeV$, $\mtau=1.777\GeV$~\cite{PDG}.  
The CP-even mixing angle is computed including the leading 
corrections up to two-loop order by means of the  
program {\tt FeynHiggsFast}~\cite{Heinemeyer:2000nz}.
The branching ratios of Higgs boson decays into $b\bar b$ and $\tau^+ \tau^-$ have been also 
computed with {\tt FeynHiggs 2.11} and we find a perfect agreement with our results for values of $\tan\beta \lesssim$ 50. The difference between these two computations for larger values of $\tan\beta$ is around $10\%$. 

\begin{figure}[t!]
\begin{center}
\begin{tabular}{cc}
\includegraphics[width=0.475\textwidth]{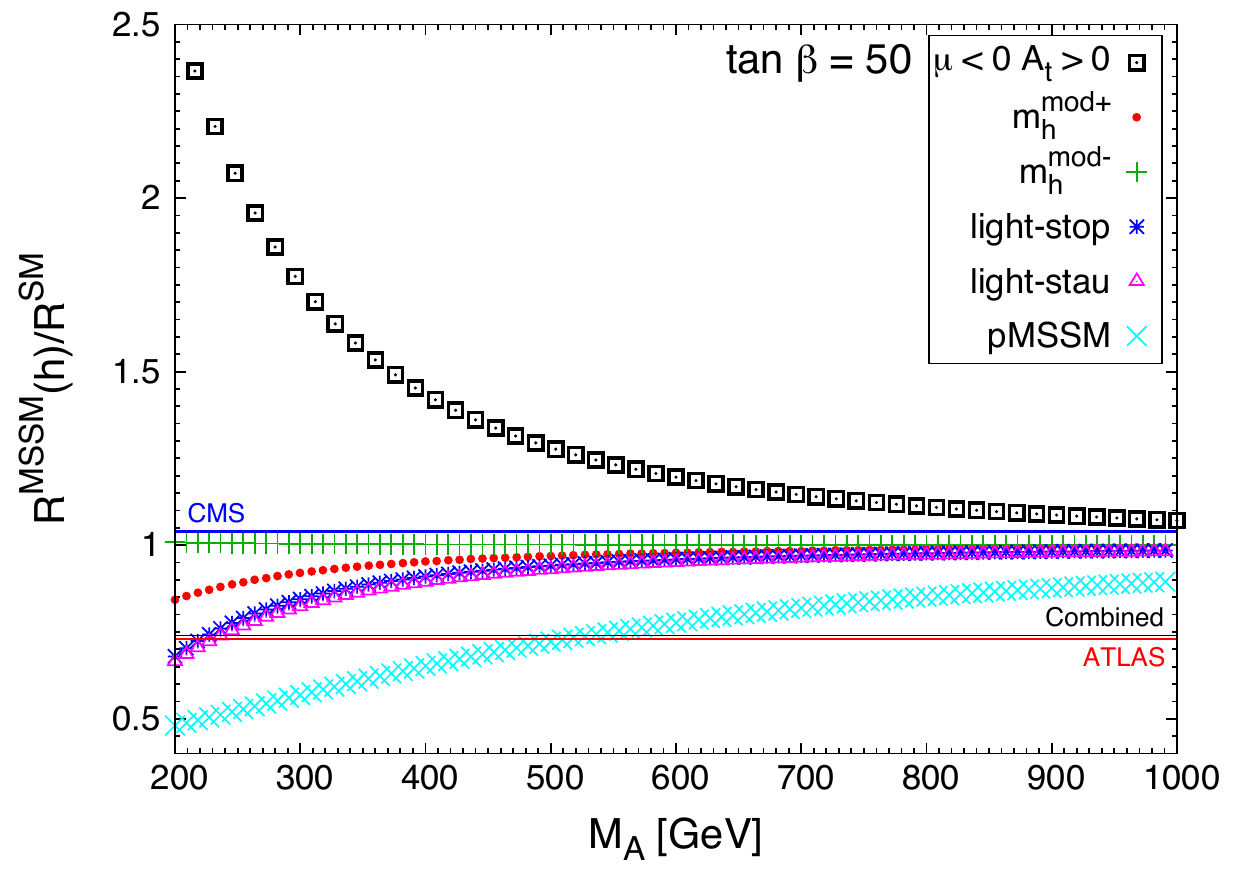}&
\includegraphics[width=0.475\textwidth]{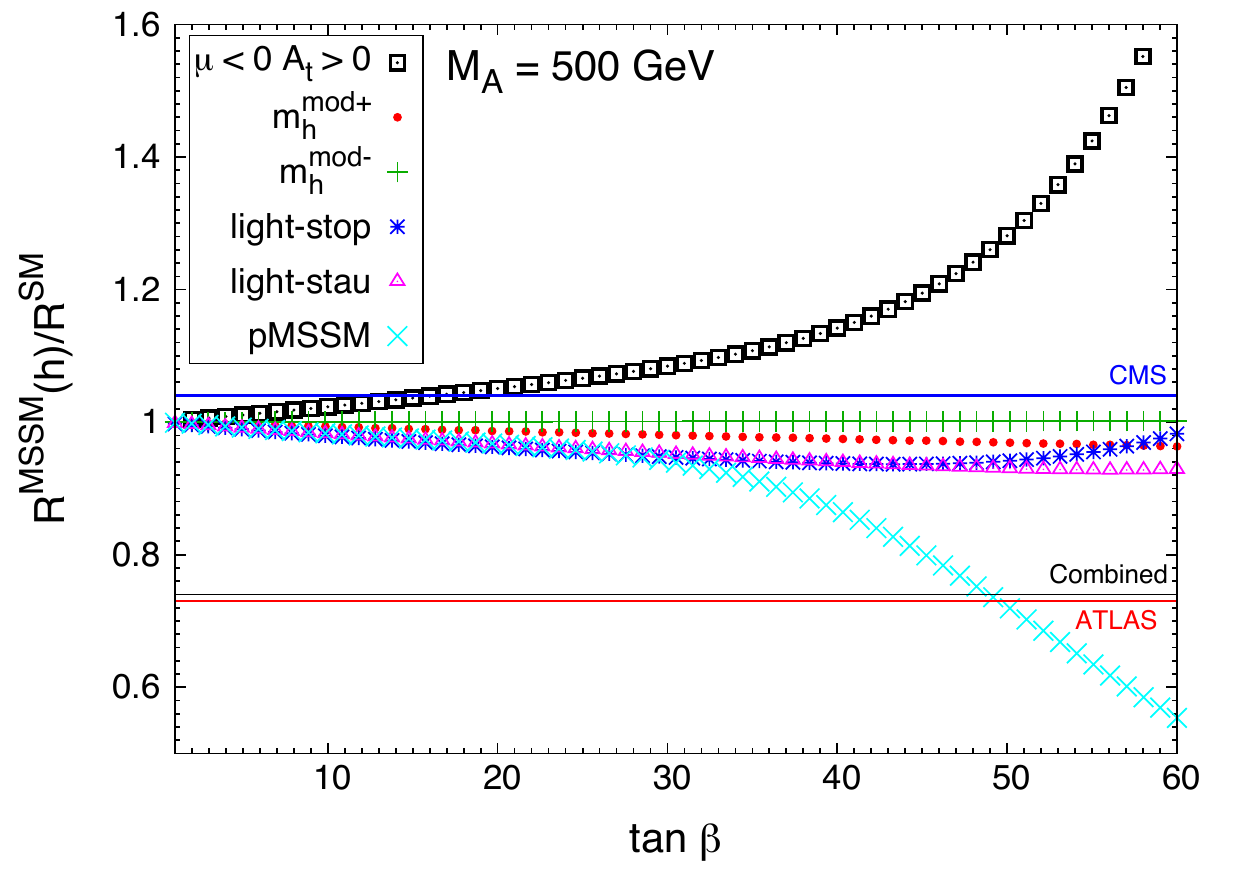}\\
(a)&(b)
\end{tabular}
\caption{Normalized ratio $X^\text{MSSM}(h)$~(\ref{eq:Rh0}), as a function of:
(a) \ma\ ($\tan\beta = 50$) and (b)  $\tan\beta$ (\ma\ $= 500$ GeV), 
for various choices of 
benchmark scenarios. In both plots, the horizontal lines show the one
standard deviation experimental upper limit regions for $X$~(\ref{eq:Xlimits}) by ATLAS (red), CMS (blue), and our combined result (black).}
\label{Rh-MAtanb}
\end{center}
\end{figure}
In Fig.~\ref{Rh-MAtanb} we present numerical results for $R^\text{MSSM}(h)$ 
normalized to the SM value, as a function of (a) \ma\ and (b)  $\tan\beta$,  
for various choices of SUSY scenarios with 
$\tan\beta = 50$ and \ma\ $= 500$ GeV,  
respectively. The horizontal lines show the 
one-standard deviation experimental upper limit for $X$~(\ref{eq:Xlimits}) 
by ATLAS (in red), CMS (in blue), and our combined
result (in black).
The largest deviation with respect to the SM value emerges in the scenario
$\mu<0$ $A_t>0$. Actually, the present analysis already excludes this
scenario at 68\% C.L., and only a small region with 
$\tb<10$ and $\ma=500\text{ GeV}$
survives the CMS measurement (Fig.~\ref{Rh-MAtanb}b). This shows the
huge potential of the observable $R$ in SUSY searches/exclusions. We
note, however, that this scenario ($\mu<0$ $A_t>0$) is also disfavored
by the constraints from the measurement of 
$BR(B_s \to \mu^+ \mu^-)$~\cite{Altmannshofer:2012ks}. 
Actually, the sign of the dominant contribution to the corrections to $R$
is proportional to $-{\rm sign}(\mu A_t)$, and since the experimental data 
on $BR(B_s \to \mu^+ \mu^-)$ disfavors $\mu A_t<0$, it selects negative 
corrections to $R$. For this reason we will not further consider the
$\mu<0$ $A_t>0$ scenario. 
The other scenarios provide a prediction for 
$R^\text{MSSM}(h)$~(\ref{eq:Rh0}) smaller
than in the SM, and then the CMS measurement alone can not exclude any
of them~(\ref{eq:Xlimits}). However, note that most scenarios have a
prediction close the SM one, and therefore the ATLAS result disfavors
them (at 68\% C.L.). The $m_h^\text{mod-}$ scenario prediction is
practically indistinguishable from the SM one, whereas the
$m_h^\text{mod+}$ has a largest deviation of a $20\%$ with respect to the
SM value, and both of them are also disfavored by ATLAS. The light-stop and
light-stau scenarios provide larger deviations, up to $40\%$ for small
$\ma$, and this small region is not disfavored by ATLAS.  The pMSSM
scenario provides larger deviations, and thus has the largest allowed
regions, for $\ma\lesssim 500\GeV$ and $\tb\gtrsim 50$.

We note in Fig.~\ref{Rh-MAtanb}b the flat evolution of the normalized
ratio $X$ with respect to $\tb$ in the $m_h^\text{mod+}$,
$m_h^\text{mod-}$, light-stop, and light-stau scenarios. The reason is
manifold: first of all, the resummation procedure softens the $\tb$
evolution; second, at $\ma\simeq500\GeV$ the MSSM Higgs sector is
already close to the decoupling limit, with $\tan\alpha$ close to
$-1/\tan\beta$ and therefore providing a small effect of the $\Delta
m_f$ corrections to $X(h)$~(\ref{eq:RA}); thirdly, those scenarios use
as input parameter in the squark sector the non-diagonal element of the
squark-mass matrix $X_{t[b,\tau]}=A_{t[b,\tau]}-\mu \cot\beta [\tan\beta]$, and
therefore the sfermion mass matrix is nearly flat as a function of
$\tb$, and so are also the sfermion masses. For the pMSSM scenario the
first two conditions also apply, however here the input parameter in the
sfermion sector is the soft-SUSY-breaking trilinear coupling
$A_{t[b,\tau]}$, therefore the sfermion-mass mixing terms changes
strongly with $\tb$, and so do also the physical sfermion masses.

As expected, the decoupling behavior with \ma\  becomes apparent in 
Fig.~\ref{Rh-MAtanb}a for all the SUSY scenarios. Notice that in all 
the above scenarios the gluino mass is around $1500 \GeV$.  We have 
also examined numerically the decoupling behavior of the ratio $X$ 
with the gluino mass, extrapolating the results up to $M_{g}\sim 5000 \GeV$. 
Our results show that there is not decoupling, the ratio $X$ tends to 
a constant value for all mentioned SUSY scenarios. Therefore, 
our conclusions are also valid for large values of the gluino mass, 
in perfect agreement with the present bounds
for this mass at the LHC.

\begin{figure}[t!]
\begin{center}
\begin{tabular}{cc}
  \includegraphics[width=0.44\textwidth]{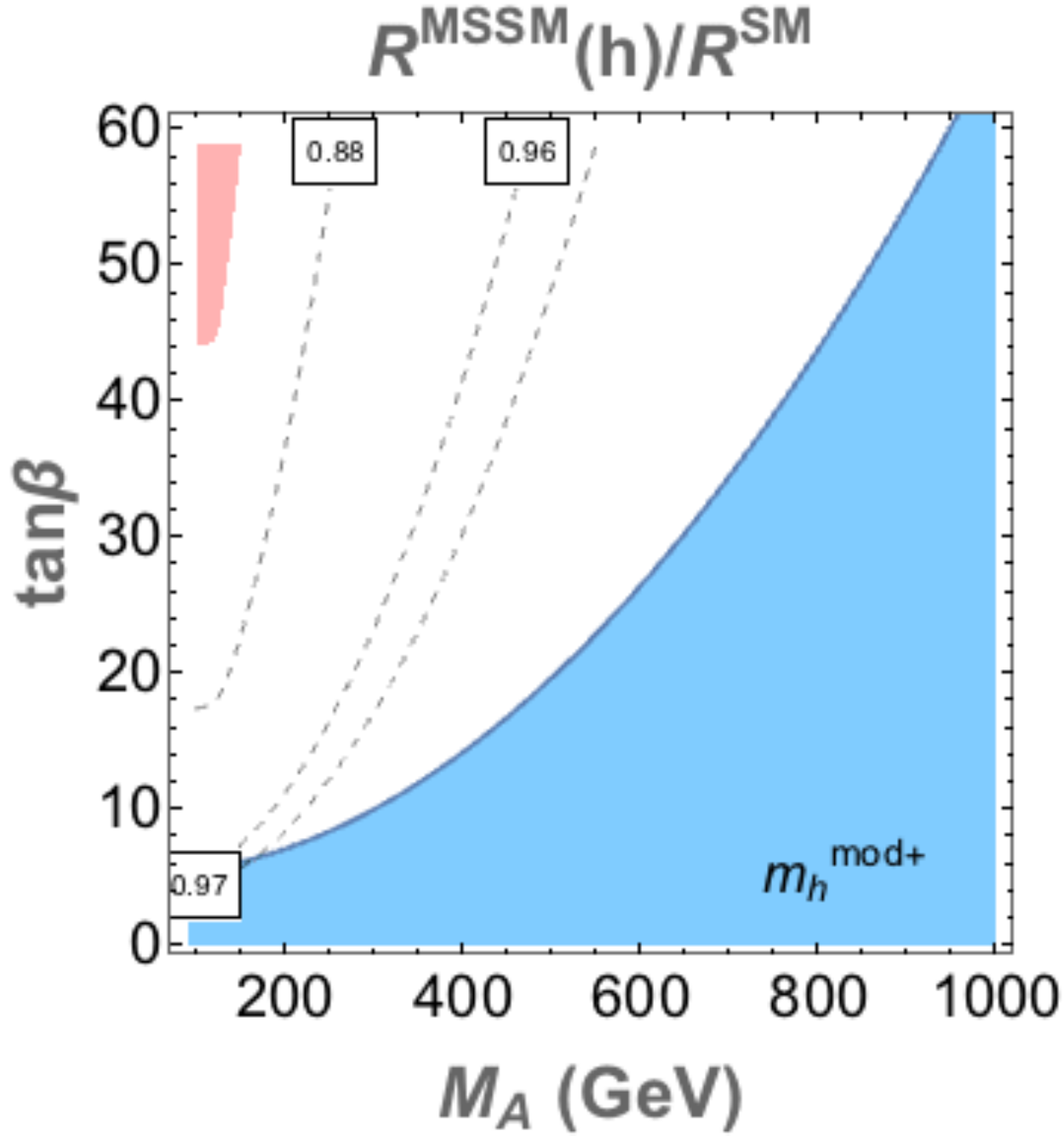} &
\includegraphics[width=0.44\textwidth]{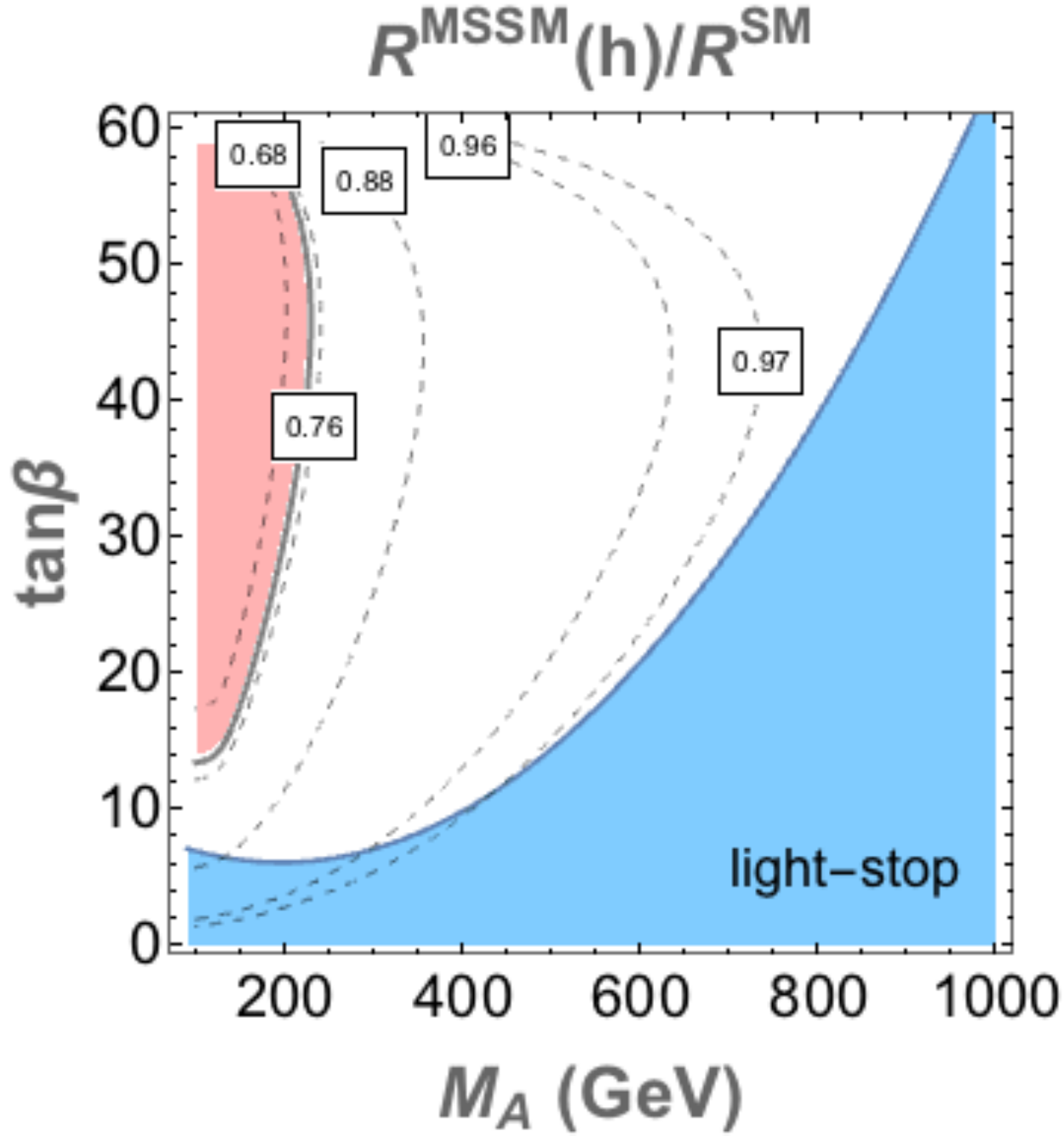}
\\
(a) & (b) \\
& \\
\includegraphics[width=0.44\textwidth]{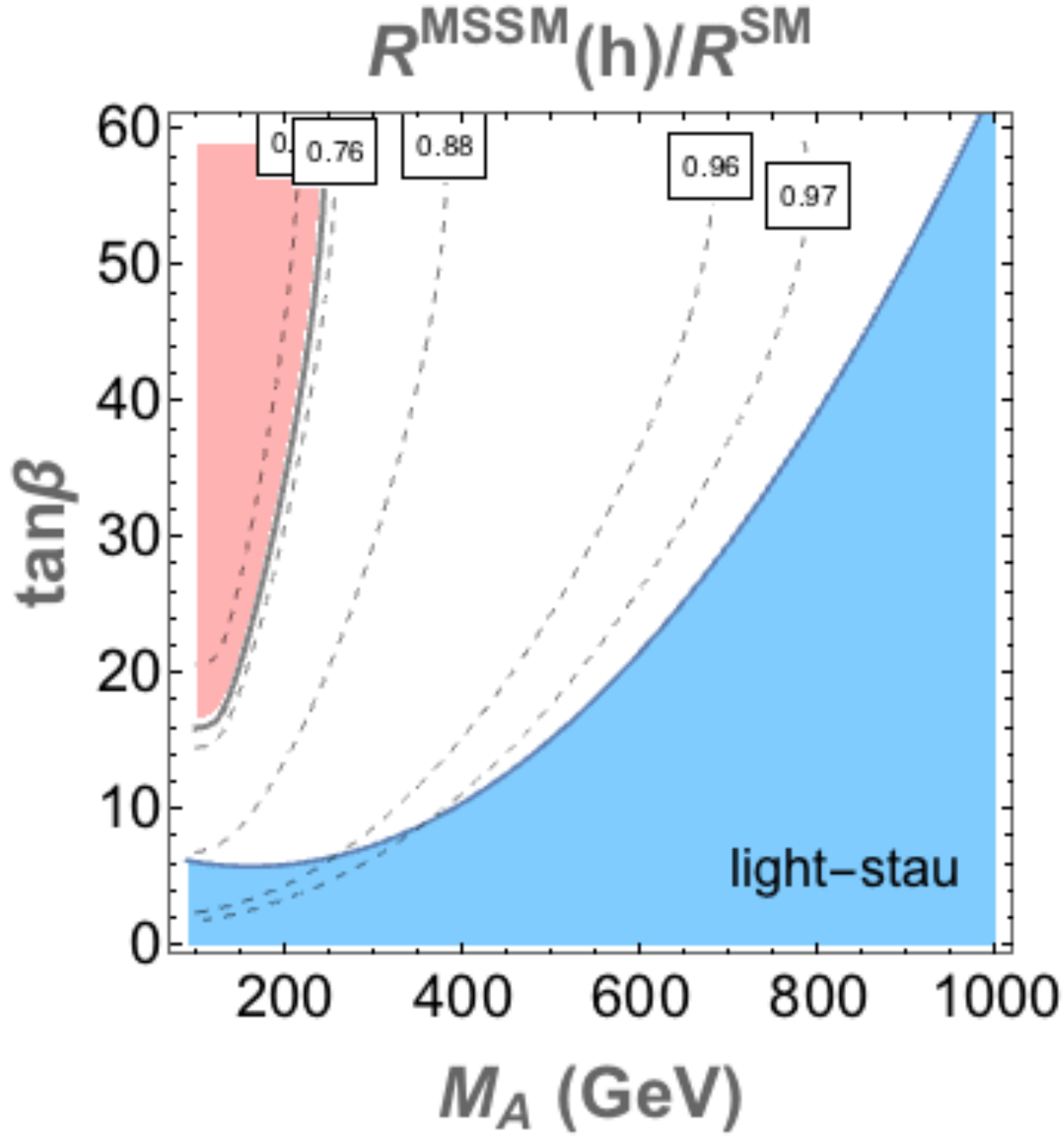} &
  \includegraphics[width=0.44\textwidth]{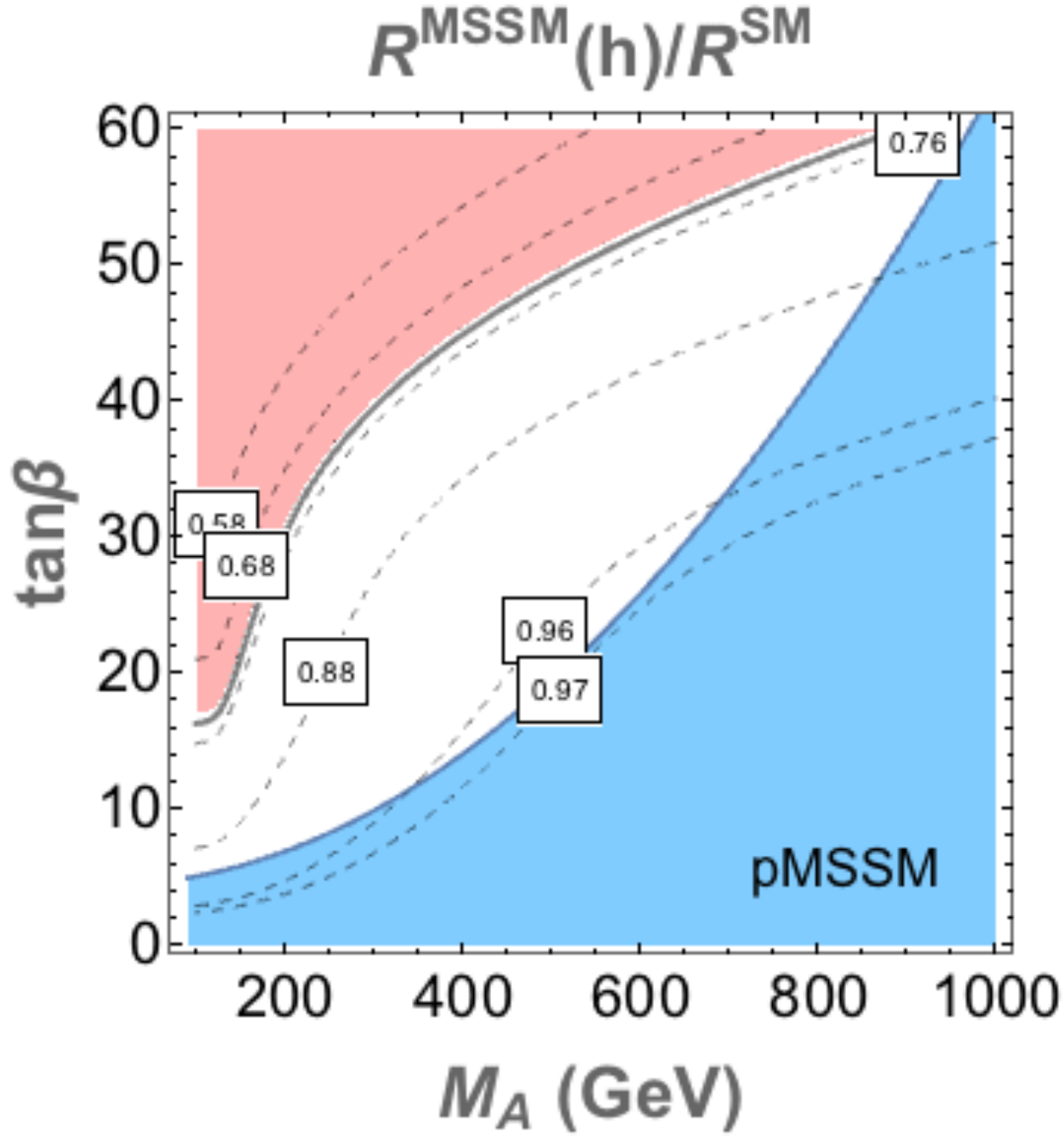}\\
(c) & (d) 
\end{tabular}
\end{center}
\caption{Contour plots in the $\ma-\tb$ plane for the normalized ratio $X$~(\ref{eq:Xlimits}), in the
 (a) $m_h^\text{mod+}$, (b) light-stop, (c) light-stau, and (d) pMSSM
 scenarios. The red [black]
  curve shows the upper (one-standard deviation) limit from
  ATLAS [our combination]~(\ref{eq:Xlimits}), the favored region is 
shown in red. Sensitivity regions on $X(h)$~(\ref{eq:Rh0}) with the
  different expected accuracies defined in Table~\ref{accuracy} are also
included. The sensitivity regions are the ones to the left of the 
corresponding curve.
Shown in blue is the 95\% C.L. allowed regions by
the negative searches by ATLAS and CMS for neutral MSSM Higgs bosons
decaying to a pair of $\tau$ leptons~\cite{MAtanb_exclusion_refs}.}
\label{fig:Xhscan}
\end{figure}
We finish this section by discussing the regions of the MSSM parameter
space favored  by the present experimental values of
$X$~(\ref{eq:Xlimits}). Of course, as already told, all the studied
scenarios have $X\lesssim1$, and therefore all of them are allowed
by CMS. Furthermore, the $m_h^\text{mod-}$ scenario has very small
deviations with respect to the SM value and it is practically indistinguishable from the SM. 
Fig.~\ref{fig:Xhscan} shows the contour plots of $X(h)$~(\ref{eq:Rh0})
in the MSSM for the  $m_h^\text{mod+}$, light-stop, light-stau and pMSSM
scenarios. The red [black] line shows the upper (one-standard deviation) limit
by ATLAS [our combination]~(\ref{eq:Xlimits}), 
the allowed region is the red area of the curve. 
We also show the 95\% C.L. favored regions (shaded blue areas) by
the negative searches by ATLAS and CMS for neutral MSSM Higgs bosons
decaying into a pair of $\tau$ leptons~\cite{MAtanb_exclusion_refs}.
We see that in the $m_h^\text{mod+}$ roughly the 
whole explored $\ma-\tb$ plane
is disfavored, whereas in the light-stop and light-stau scenarios a small
corner of large $\tb$ and low $\ma$ is favored. The region favored in
the pMSSM scenario is much larger, allowing large values of $\ma$
with large $\tb$. In all the cases, the favored regions fall completely
inside the excluded region for the CMS and ATLAS direct searches for
Higgs bosons decaying into $\tau$-lepton pairs, which means that there is a
tension (albeit a very soft one) between the experimental determination
of the Higgs boson couplings and the direct search for Higgs boson
decaying into $\tau$-lepton pairs.

\section{Future prospects}
\label{sensitivities}

\begin{table}
\begin{center}
\begin{tabular}{|c||c|c||c||c|}
\hline
Observable & LHC & HL-LHC & LC & HL-LHC+LC \\ \hline \hline
$H b \bar b$ & 10-13\% & 4-7\% & 0.6\% & 0.6\% \\ \hline
$H \tau^+ \tau^-$ & 6-8\% & 2-5\% & 1.3\% & 1.2\% \\ \hline \hline
$R$ & 32-42\% & 12-24\% & 4\% & 3\% \\ \hline
\end{tabular}
\caption{Expected accuracies for the measurements of the Higgs boson couplings $H b \bar b$
  and $H \tau^+ \tau^-$~\cite{accuracy-refs} and the ratio $R$~(\ref{eq:Rdef}) at the LHC/HL-LHC, LC, and in combined analyses of the HL-LHC and LC.}
\label{accuracy}
\end{center}
\end{table}

In this section we study the prospects for finding deviations in the
ratio $R$~(\ref{eq:Rdef}) in future colliders.  In order to 
define the different sensitivity regions we show in Table~\ref{accuracy} the expected accuracies 
with which the fundamental Higgs couplings $H b \bar b$ and $H \tau^+ \tau^-$, and our derived 
observable $R$~(\ref{eq:Rdef}), can be measured at the 
LHC/HL-LHC, the LC, and in combined analyses of the HL-LHC and the
LC~\cite{accuracy-refs}. Note that Table~\ref{accuracy} shows the
accuracy expected on absolute coupling measurements, whereas for the
purpose of the present work relative coupling measurements, like the
ones on eq.~(\ref{eq:lambdasexps}), are sufficient, and those have
better accuracies.

We reanalyze, from the point of view of the sensitivity to the SUSY
nature of the neutral MSSM Higgs bosons, the results of
Fig.~\ref{fig:Xhscan}, 
where the regions in the $\ma-\tb$ plane in which the MSSM
prediction for the normalized ratio $X$~(\ref{eq:Xnorm}) is larger than
the expected sensitivities of Table~\ref{accuracy} are depicted. 
Fig.~\ref{fig:Xhscan} shows these sensitivity regions on
$X(h)$~(\ref{eq:Rh0}) for
the $m_h^\text{mod+}$, light-stop, light-stau, and pMSSM scenarios,
respectively, for $42\%, 32\%, 24\%, 12\%, 4\%$, and $3\%$ accuracy
measurements. The sensitive regions are the ones above and to the left of
the corresponding curve.
The sensitivity regions for the $m_h^\text{mod-}$ scenario are not shown
here, since as can be inferred from the results of Fig.~\ref{Rh-MAtanb},
it is not possible to distinguish its predictions from the SM
ones. Indeed, in order to measure a deviation with respect the SM value
in this scenario, an accuracy of at least $\sim$ 0.5\% would be required.

The SUSY nature of the discovered Higgs boson of 125 GeV is 
testable within these four scenarios with the expected accuracies for
the current LHC runs or for its high luminosity phase. Unfortunately 
the corresponding sensitivity regions lie mainly outside the shaded
blue areas and thus are excluded by the ATLAS and CMS direct searches. 
Only in the pMSSM scenario with a 12\%
measurement (corresponding to the HL-LHC accuracy) one can have
sensitivity to SUSY in a favored area for large values of 
$M_A$ (around 800-1000 GeV) and
$\tan\beta \gtrsim$ 50. If we turn to the LC and combined HL-LHC+LC
accuracies, the possibility of detecting a deviation with respect to the
SM value becomes more favored. In that case, within the
$m_h^\text{mod+}$ scenario, one could be sensitive to SUSY in the region
with very low values of $\tan\beta$ and $M_A$, up to $M_A \sim$ 200
GeV. On the other hand, within the light-stop and light-stau scenarios
the LC sensitivities are kept up to $M_A \simeq$ 400 GeV. From this
value of $M_A$, the sensitivity regions lie in the area of exclusion and
are not allowed. The contour lines for the LC accuracies in
the pMSSM scenario are allowed for any value of $M_A$,
depending on the value of $\tan\beta$. Then, if this class of scenario
is realized in nature, one would be able to observe deviations with
respect the SM predictions at a possible future LC that would mean a
clear hint of SUSY. 

\begin{figure}[t!]
\begin{center}
\begin{tabular}{cc}
\includegraphics[width=0.60\textwidth]{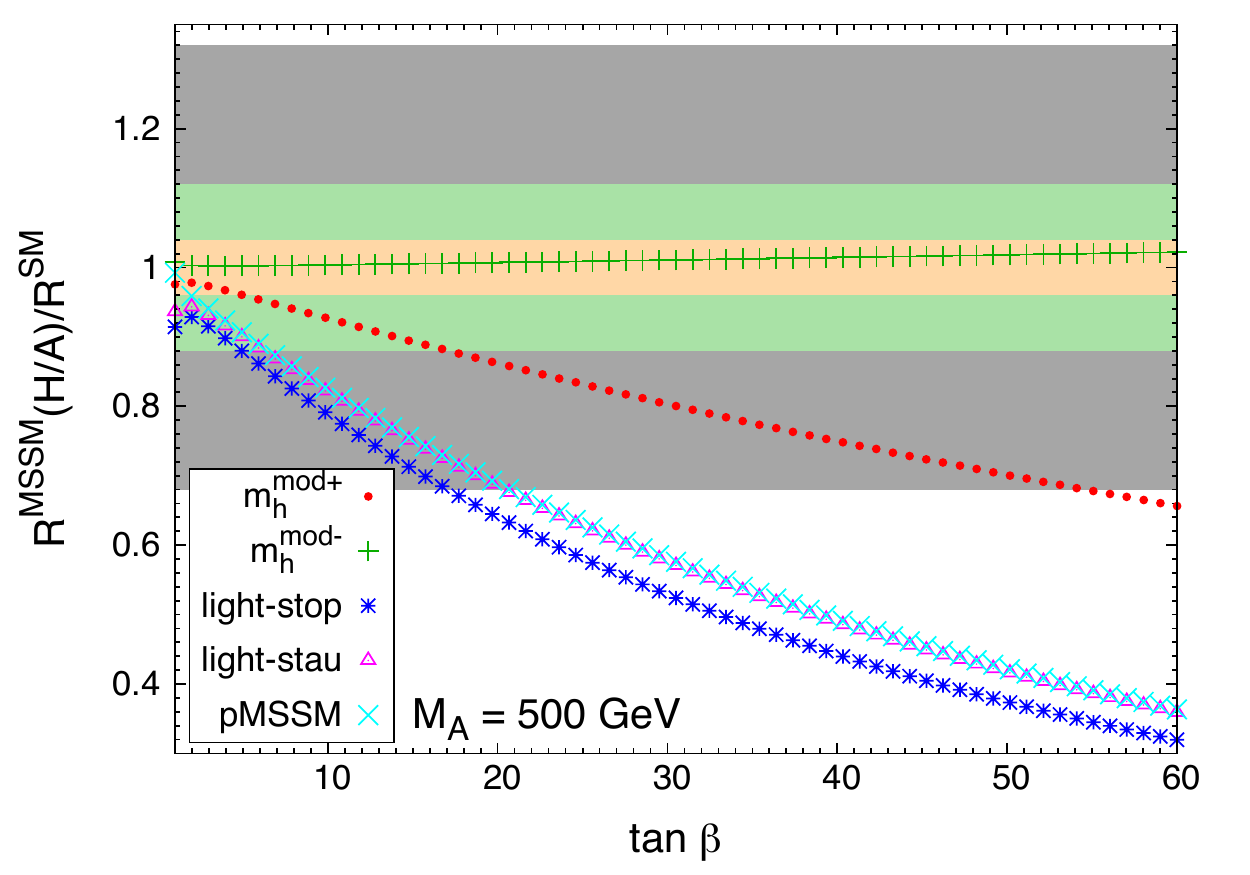}
\end{tabular}
\caption{Deviation of $R^\text{MSSM}(H/A)$ with respect to the SM value, 
as a function of $\tan\beta$ for various choices of benchmark scenarios. 
The shaded gray region shows the $\pm$32\% deviation whit respect to the SM, 
the shaded green one the $\pm$12\%, and the shaded orange one the $\pm$4\%.}
\label{RHA-tanb}
\end{center}
\end{figure}
We turn now our attention to the heavy neutral Higgs bosons $H$ and $A$. In
case of these heavy states are found at the LHC, one still has to answer
the question whether they belong to a simple 2HDM, or whether they
belong to a SUSY extension of the SM. The ratio of branching ratios
$R$~(\ref{eq:Rdef}) can be useful in this task. Fig.~\ref{RHA-tanb}
shows the normalized ratio $X(H/A)$~(\ref{eq:Xnorm}) as a function of
$\tb$ for the $m_h^\text{mod+}$, light-stop, light-stau, 
and pMSSM scenarios with $\ma=500\GeV$. 
Note that once we are close to the decoupling limit
($\ma\gg\mz$) the couplings of $H/A$ are indistinguishable, and,
furthermore, the ratio $R(H/A)$ becomes independent of $\ma$. We show through
different shaded regions the expected accuracies for the future measurement of
$R$~(\ref{eq:Rdef}), $\pm$32\% (shaded gray
area), $\pm$12\% (shaded green area), and $\pm$4\% (shaded orange area).  
For the sake of readiness, we only show the smallest
accuracies reported in Table~\ref{accuracy}. 
We see that, given a large enough value of $\tb$, all the scenarios (except
the $m_h^\text{mod-}$) provide a value for $X(H/A)$~(\ref{eq:Xnorm})
larger than the expected experimental accuracies. 
Within the $m_h^\text{mod+}$ scenario, it would be possible to
observe at the LHC 32\% deviations with respect to the SM value for
$\tan\beta \gtrsim 55$. At the HL-LHC, we could be sensitive to 
SUSY within this scenario for a 12\% deviation from values of
$\tan\beta \gtrsim$ 20. 
The results for the light-stop, light-stau, and
pMSSM scenarios are very similar and even more favorable in order to
detect any SUSY deviation with respect to the SM value. The LHC could
observe $32\%$ deviations for values of
$\tan\beta$ larger than 20 and the HL-LHC would be sensitive
to SUSY with $12\%$ deviations for $\tan\beta \gtrsim$ 5. 
If an accuracy of $4\%$ is achieved at a future LC, it would be 
possible to probe the SUSY nature of $H$ and $A$
Higgs bosons for $\tan\beta \gtrsim$
5 in any of these four scenarios. Therefore, if a new heavy Higgs 
scalar or pseudoscalar is discovered,
and its couplings to bottom quarks and $\tau$ leptons are measured with
a moderate level of precision, it would be possible to distinguish
between SUSY and non-SUSY Higgs sectors at the LHC. 
 
\section{Conclusions}
\label{conclusions}

In this work, we have updated the analysis of the observable 
$R = \text{BR}(H \to b \bar b)/\text{BR}(H \to \tau^+ \tau^-)$~(\ref{eq:Rdef}) in order
to look for a strong evidence for, or against, the SUSY nature of the
Higgs boson. We have considered more realistic MSSM scenarios with a
lightest Higgs boson mass $m_h$ compatible with the current value of the 
Higgs boson mass $m_{H^\text{SM}} \simeq$ 125 GeV. 
We have compared the theoretical prediction in the MSSM with the current
experimental determination of Higgs boson couplings to fermions at the
LHC~(\ref{eq:lambdasexps}). We find that the SM prediction for $R$ agrees well
with current CMS data, but using ATLAS data we obtain a (one-standard
deviation) upper limit below the SM prediction. By contrast, 
the SUSY contributions can
provide a prediction that agrees with the experiment at the 
one-standard deviation level. 
Current accuracy already allows to exclude portions of the parameter
space, showing the potential of the observable $R$ to discriminate among
different models of new physics.
It is also important to mention that the parameter space regions that 
are favored by the determination of the
Higgs boson couplings to fermions are in tension with the
direct searches for MSSM neutral Higgs boson decaying into $\tau$-lepton pairs.

We have also looked at the prospects for future measurements of the
Higgs boson couplings. We find that, in wide regions of the parameter
space, a moderate accuracy of the couplings would signal the presence of
SUSY in the Higgs boson data. This analysis allows the use of relative
couplings, which can improve significantly the accuracy in the
experimental determination of $R$.

Finally, we have moved our attention to the heavier Higgs bosons 
of the MSSM, $H$ and $A$. If one or both of these heavy neutral Higgs bosons 
are discovered, one would still need to determine whether
they belong to a generic 2HDM or to a SUSY model. A moderate accuracy
determination of their couplings to $b$ quarks and $\tau$ leptons, 
by means of the analysis of the ratio $R$, would be sufficient 
to discern the SUSY nature of such particles.

\section*{Acknowledgments}

The work of E.~A. is funded by a
grant from the Spanish Consolider-Ingenio 2010 Program 
CPAN (CSD2007-00042). 
E.~A. and S.~P. are financially supported by the Spanish DGIID-DGA 
Grant No. 2013-E24/2, the Spanish MINECO Grants No. FPA2012-35453 
and CPAN (CSD2007-00042). 
J.G. has been supported by MINECO (Spain) (FPA2013-46570-C2-2-P),
by DURSI (2014-SGR-1474) and CPAN (CSD2007-00042). 

\bibliographystyle{unsrt}

\begin{thebibliography}{99}

\bibitem{HiggsDiscovery}
%\bibitem{Aad:2012tfa}
  G.~Aad {\it et al.}  [ATLAS Collaboration],
  Phys.\ Lett.\ B {\bf 716} (2012) 1
  [arXiv:1207.7214 [hep-ex]];
  %%CITATION = ARXIV:1207.7214;%%
%\bibitem{Chatrchyan:2012ufa}
  S.~Chatrchyan {\it et al.}  [CMS Collaboration],
  Phys.\ Lett.\ B {\bf 716} (2012) 30
  [arXiv:1207.7235 [hep-ex]].
  %%CITATION = ARXIV:1207.7235;%%
  
%\cite{Aad:2015zhl}
\bibitem{Aad:2015zhl}
  G.~Aad {\it et al.}  [ATLAS and CMS Collaborations],
  %``Combined Measurement of the Higgs Boson Mass in $pp$ Collisions at $\sqrt{s}=7$ and 8 TeV with the ATLAS and CMS Experiments,''
  Phys.\ Rev.\ Lett.\  {\bf 114} (2015) 191803
  [arXiv:1503.07589 [hep-ex]].
  %%CITATION = ARXIV:1503.07589;%%
  %46 citations counted in INSPIRE as of 26 May 2015

\bibitem{Planck}
%\bibitem{Altarelli:1994rb}
  G.~Altarelli and G.~Isidori,
  %``Lower limit on the Higgs mass in the standard model: An Update,''
  Phys.\ Lett.\ B {\bf 337} (1994) 141;
  %%CITATION = PHLTA,B337,141;%%
  %\bibitem{Casas:1994qy}
  J.~A.~Casas, J.~R.~Espinosa and M.~Quiros,
  %``Improved Higgs mass stability bound in the standard model and implications for supersymmetry,''
  Phys.\ Lett.\ B {\bf 342} (1995) 171 [hep-ph/9409458];
  %%CITATION = HEP-PH/9409458;%%
  %\bibitem{Casas:1996aq}
  %J.~A.~Casas, J.~R.~Espinosa and M.~Quiros,
  %``Standard model stability bounds for new physics within LHC reach,''
  Phys.\ Lett.\ B {\bf 382} (1996) 374
  [hep-ph/9603227];
  %%CITATION = HEP-PH/9603227;%%
  %\bibitem{Hambye:1996wb}
  T.~Hambye and K.~Riesselmann,
  %``Matching conditions and Higgs mass upper bounds revisited,''
  Phys.\ Rev.\ D {\bf 55} (1997) 7255
  [hep-ph/9610272];
  %%CITATION = HEP-PH/9610272;%%
%\bibitem{Ellis:2009tp}
  J.~Ellis {\it et al.}, 
%J.~R.~Espinosa, G.~F.~Giudice, A.~Hoecker and A.~Riotto,
  %``The Probable Fate of the Standard Model,''
  Phys.\ Lett.\ B {\bf 679} (2009) 369 [arXiv:0906.0954 [hep-ph]];
  %%CITATION = ARXIV:0906.0954;%%
  %132 citations counted in INSPIRE as of 23 Jun 2015
%\bibitem{Degrassi:2012ry}
  G.~Degrassi {\it et al.}, 
%S.~Di Vita, J.~Elias-Miro, J.~R.~Espinosa, G.~F.~Giudice, G.~Isidori and A.~Strumia,
  %``Higgs mass and vacuum stability in the Standard Model at NNLO,''
  JHEP {\bf 1208} (2012) 098 [arXiv:1205.6497 [hep-ph]];
  %%CITATION = ARXIV:1205.6497;%%
  %467 citations counted in INSPIRE as of 23 juin 2015
%\bibitem{Buttazzo:2013uya}
  D.~Buttazzo {\it et al.}, 
%G.~Degrassi, P.~P.~Giardino, G.~F.~Giudice, F.~Sala, A.~Salvio and A.~Strumia,
  %``Investigating the near-criticality of the Higgs boson,''
  JHEP {\bf 1312} (2013) 089
  [arXiv:1307.3536 [hep-ph]].
  %%CITATION = ARXIV:1307.3536;%%
  %303 citations counted in INSPIRE as of 23 juin 2015

\bibitem{Hunter}  
J.F. Gunion, H.E. Haber, G.L. Kane, S. Dawson, \textit{The 
Higgs Hunters' Guide} (Addison-Wesley, Menlo-Park, 1990). 

\bibitem{Gunion:2002zf}
  J.~F.~Gunion and H.~E.~Haber,
  %``The CP conserving two Higgs doublet model: The Approach to the decoupling limit,''
  Phys.\ Rev.\ D {\bf 67} (2003) 075019
  [hep-ph/0207010].
  %%CITATION = HEP-PH/0207010;%%
  %306 citations counted in INSPIRE as of 23 juin 2015

\bibitem{MSSM}
%\bibitem{Nilles:1983ge}
  H.~P.~Nilles,
  %``Supersymmetry, Supergravity and Particle Physics,''
  Phys.\ Rept.\  {\bf 110} (1984) 1;
  %%CITATION = PRPLC,110,1;%%
  %4475 citations counted in INSPIRE as of 02 Jun 2015
%\bibitem{Haber:1984rc}
  H.~E.~Haber and G.~L.~Kane,
  %``The Search for Supersymmetry: Probing Physics Beyond the Standard Model,''
  Phys.\ Rept.\  {\bf 117} (1985) 75;
  %%CITATION = PRPLC,117,75;%%
  %4329 citations counted in INSPIRE as of 02 Jun 2015
%\bibitem{Lahanas:1986uc}
  A.~B.~Lahanas and D.~V.~Nanopoulos,
  %``The Road to No Scale Supergravity,''
  Phys.\ Rept.\  {\bf 145} (1987) 1.
  %%CITATION = PRPLC,145,1;%%
  %729 citations counted in INSPIRE as of 02 Jun 2015

%\cite{Guasch:2001wv}
\bibitem{Guasch:2001wv}
  J.~Guasch, W.~Hollik and S.~Pe\~naranda,
  %``Distinguishing Higgs models in H ---> b anti-b / H ---> tau+ tau-,''
  Phys.\ Lett.\ B {\bf 515} (2001) 367
  [hep-ph/0106027].
  %%CITATION = HEP-PH/0106027;%%
  %74 citations counted in INSPIRE as of 26 May 2015
  
%%%%%%%%%%%%%%%%%%%%%%%%%%%%%%%
% DeltaMb references

%\cite{Hall:1993gn}
\bibitem{Hall:1993gn}
  L.~J.~Hall, R.~Rattazzi and U.~Sarid,
  %``The Top quark mass in supersymmetric SO(10) unification,''
  Phys.\ Rev.\ D {\bf 50} (1994) 7048
  [hep-ph/9306309, hep-ph/9306309].
  %%CITATION = HEP-PH/9306309,;%%
  %808 citations counted in INSPIRE as of 29 May 2015

%\cite{Carena:1994bv}
\bibitem{Carena:1994bv}
  M.~Carena, M.~Olechowski, S.~Pokorski and C.~E.~M.~Wagner,
  %``Electroweak symmetry breaking and bottom - top Yukawa unification,''
  Nucl.\ Phys.\ B {\bf 426} (1994) 269
  [hep-ph/9402253].
  %%CITATION = HEP-PH/9402253;%%
  %711 citations counted in INSPIRE as of 29 May 2015
 
%\cite{Carena:1999py}
\bibitem{Carena:1999py}
  M.~Carena, D.~Garcia, U.~Nierste and C.~E.~M.~Wagner,
  %``Effective Lagrangian for the $\bar{t} b H^{+}$ interaction in the MSSM and charged Higgs phenomenology,''
  Nucl.\ Phys.\ B {\bf 577} (2000) 88
  [hep-ph/9912516].
  %%CITATION = HEP-PH/9912516;%%
  %499 citations counted in INSPIRE as of 29 May 2015

%\cite{Guasch:2003cv}
\bibitem{Guasch:2003cv}
  J.~Guasch, P.~Hafliger and M.~Spira,
  %``MSSM Higgs decays to bottom quark pairs revisited,''
  Phys.\ Rev.\ D {\bf 68} (2003) 115001
  [hep-ph/0305101].
  %%CITATION = HEP-PH/0305101;%%
  %111 citations counted in INSPIRE as of 01 Jun 2015

\bibitem{2loop-resummation}  
%\cite{Noth:2010jy}
%\bibitem{Noth:2010jy}
  D.~Noth and M.~Spira,
  %``Supersymmetric Higgs Yukawa Couplings to Bottom Quarks at next-to-next-to-leading Order,''
  JHEP {\bf 1106} (2011) 084
  [arXiv:1001.1935 [hep-ph]];
  %%CITATION = ARXIV:1001.1935;%%
  %\bibitem{Crivellin:2012zz}
  A.~Crivellin and C.~Greub,
  %``Two-loop supersymmetric QCD corrections to Higgs-quark-quark couplings in the generic MSSM,''
  Phys.\ Rev.\ D {\bf 87} (2013) 015013
   [Phys.\ Rev.\ D {\bf 87} (2013) 079901]
  [arXiv:1210.7453 [hep-ph]] and references therein.
  %%CITATION = ARXIV:1210.7453;%%
  %17 citations counted in INSPIRE as of 01 Jul 2015

%%%%%%%%%%%%%%%%%%%%%%%%%%%%%%%%%%%%%%%%%%%%%%%%%%%%%%%%%%%%%%%%%%%%%%%%%%%
%\cite{Dabelstein:1995js}
\bibitem{Dabelstein:1995js}
  A.~Dabelstein,
  %``Fermionic decays of neutral MSSM Higgs bosons at the one loop level,''
  Nucl.\ Phys.\ B {\bf 456} (1995) 25
  [hep-ph/9503443].
  %%CITATION = HEP-PH/9503443;%%
  %195 citations counted in INSPIRE as of 29 May 2015
  
%\cite{Coarasa:1995yg}
\bibitem{Coarasa:1995yg}
  J.~A.~Coarasa Perez, R.~A.~Jimenez and J.~Sola,
  %``Strong effects on the hadronic widths of the neutral Higgs bosons in the MSSM,''
  Phys.\ Lett.\ B {\bf 389} (1996) 312
  [hep-ph/9511402].
  %%CITATION = HEP-PH/9511402;%%
  %99 citations counted in INSPIRE as of 29 May 2015
  
%\cite{Pierce:1996zz}
\bibitem{Pierce:1996zz}
  D.~M.~Pierce, J.~A.~Bagger, K.~T.~Matchev and R.~j.~Zhang,
  %``Precision corrections in the minimal supersymmetric standard model,''
  Nucl.\ Phys.\ B {\bf 491} (1997) 3
  [hep-ph/9606211].
  %%CITATION = HEP-PH/9606211;%%
  %878 citations counted in INSPIRE as of 29 May 2015
  
%\cite{Coarasa:1996qa}
\bibitem{Coarasa:1996qa}
  J.~A.~Coarasa Perez  {\it et al.}, 
%D.~Garcia, J.~Guasch, R.~A.~Jimenez and J.~Sola,
  %``Quantum effects on $t \to H$ + $b$ in the MSSM: A Window to ' virtual ' supersymmetry?,''
  Eur.\ Phys.\ J.\ C {\bf 2} (1998) 373
  [hep-ph/9607485].
  %%CITATION = HEP-PH/9607485;%%
  %105 citations counted in INSPIRE as of 29 May 2015
  
%\cite{Carena:1998gk}
\bibitem{Carena:1998gk}
  M.~Carena, S.~Mrenna and C.~E.~M.~Wagner,
  %``MSSM Higgs boson phenomenology at the Tevatron collider,''
  Phys.\ Rev.\ D {\bf 60} (1999) 075010
  [hep-ph/9808312].
  %%CITATION = HEP-PH/9808312;%%
  %249 citations counted in INSPIRE as of 29 May 2015

%\cite{Carena:1999bh}
\bibitem{Carena:1999bh}
  M.~Carena, S.~Mrenna and C.~E.~M.~Wagner,
  %``The Complementarity of LEP, the Tevatron and the CERN LHC in the search for a light MSSM Higgs boson,''
  Phys.\ Rev.\ D {\bf 62} (2000) 055008
  [hep-ph/9907422].
  %%CITATION = HEP-PH/9907422;%%
  %144 citations counted in INSPIRE as of 29 May 2015
     
%\cite{Babu:1998er}
\bibitem{Babu:1998er}
  K.~S.~Babu and C.~F.~Kolda,
  %``Signatures of supersymmetry and Yukawa unification in Higgs decays,''
  Phys.\ Lett.\ B {\bf 451} (1999) 77
  [hep-ph/9811308].
  %%CITATION = HEP-PH/9811308;%%
  %90 citations counted in INSPIRE as of 29 May 2015  

%\cite{Haber:2000kq}
\bibitem{Haber:2000kq}
  H.~E.~Haber {\it et al.}, 
%M.~J.~Herrero, H.~E.~Logan, S.~Pe\~naranda, S.~Rigolin and D.~Temes,
  %``SUSY QCD corrections to the MSSM h0 $b \bar{b}$ vertex in the decoupling limit,''
  Phys.\ Rev.\ D {\bf 63} (2001) 055004
  [hep-ph/0007006].
  %%CITATION = HEP-PH/0007006;%%
  %97 citations counted in INSPIRE as of 29 May 2015

%\cite{Heinemeyer:2000fa}
\bibitem{Heinemeyer:2000fa}
  S.~Heinemeyer, W.~Hollik and G.~Weiglein,
  %``Decay widths of the neutral CP even MSSM Higgs bosons in the Feynman diagrammatic approach,''
  Eur.\ Phys.\ J.\ C {\bf 16} (2000) 139
  [hep-ph/0003022].
  %%CITATION = HEP-PH/0003022;%%
  %100 citations counted in INSPIRE as of 29 May 2015

%\cite{Heinemeyer:2001pq}
\bibitem{Heinemeyer:2001pq}
  S.~Heinemeyer and G.~Weiglein,
  %``Higgs production and decay in the MSSM: Status and perspectives,''
  AIP Conf.\ Proc.\  {\bf 578} (2001) 275
  [hep-ph/0102117].
  %%CITATION = HEP-PH/0102117;%%
  %10 citations counted in INSPIRE as of 03 Jun 2015

\bibitem{CDHPTdeTr}
%\cite{Herrero:2001yg}
%\bibitem{Herrero:2001yg}
  M.~J.~Herrero, S.~Pe\~naranda and D.~Temes,
  %``SUSY - QCD decoupling properties in $H^{+} \to t \bar{b}$ decay,''
  Phys.\ Rev.\ D {\bf 64} (2001) 115003
  [hep-ph/0105097];
  %%CITATION = HEP-PH/0105097;%%
  %\bibitem{Dobado:2001mq}
  A.~Dobado, M.~J.~Herrero and D.~Temes,
  %``Effective Higgs quark quark couplings from a heavy SUSY spectrum,''
  Phys.\ Rev.\ D {\bf 65} (2002) 075023
  [hep-ph/0107147];
  %%CITATION = HEP-PH/0107147;%%
  %\bibitem{Curiel:2001ad}
  A.~M.~Curiel  {\it et al.}, 
%M.~J.~Herrero, D.~Temes and J.~F.~De Troconiz,
  %``Optimal observables to search for indirect SUSY QCD signals in Higgs bosons decays,''
  Phys.\ Rev.\ D {\bf 65} (2002) 075006
  [hep-ph/0106267].
  %%CITATION = HEP-PH/0106267;%%
  %13 citations counted in INSPIRE as of 03 Jun 2015
  
\bibitem{higgsradcor}
%\cite{Carena:2000dp}
%\bibitem{Carena:2000dp}
  M.~Carena  {\it et al.}, 
%H.~E.~Haber, S.~Heinemeyer, W.~Hollik, C.~E.~M.~Wagner and G.~Weiglein,
  %``Reconciling the two loop diagrammatic and effective field theory computations of the mass of the lightest CP - even Higgs boson in the MSSM,''
  Nucl.\ Phys.\ B {\bf 580} (2000) 29
  [hep-ph/0001002];
  %%CITATION = HEP-PH/0001002;%%
  %\bibitem{Espinosa:2000df}
  J.~R.~Espinosa and R.~J.~Zhang,
  %``Complete two loop dominant corrections to the mass of the lightest CP even Higgs boson in the minimal supersymmetric standard model,''
  Nucl.\ Phys.\ B {\bf 586} (2000) 3
  [hep-ph/0003246];
  %\bibitem{Kitahara:2013lfa}
  T.~Kitahara and T.~Yoshinaga,
  %``Stau with Large Mass Difference and Enhancement of the Higgs to Diphoton Decay Rate in the MSSM,''
  JHEP {\bf 1305} (2013) 035
  [arXiv:1303.0461 [hep-ph]];
  %%CITATION = ARXIV:1303.0461;%%
  %\bibitem{Carena:2013iba}
  M.~Carena {\it et al.}, 
%S.~Gori, N.~R.~Shah, C.~E.~M.~Wagner and L.~T.~Wang,
  %``Light Stops, Light Staus and the 125 GeV Higgs,''
  JHEP {\bf 1308} (2013) 087 [arXiv:1303.4414 [hep-ph]];
  %%CITATION = ARXIV:1303.4414,;%%
   %\bibitem{Carena:2013ooa}
  M.~Carena {\it et al.}, 
%I.~Low, N.~R.~Shah and C.~E.~M.~Wagner,
  %``Impersonating the Standard Model Higgs Boson: Alignment without Decoupling,''
  JHEP {\bf 1404} (2014) 015
  [arXiv:1310.2248 [hep-ph]];
  %%CITATION = ARXIV:1310.2248;%%
  %\bibitem{Delgado:2013gza}
  A.~Delgado, M.~Garcia and M.~Quiros,
  %``Electroweak and supersymmetry breaking from the Higgs boson discovery,''
  Phys.\ Rev.\ D {\bf 90} (2014) 1,  015016
  [arXiv:1312.3235 [hep-ph]];
  %%CITATION = ARXIV:1312.3235;%%
  %\bibitem{Draper:2013oza}
  P.~Draper, G.~Lee and C.~E.~M.~Wagner,
  %``Precise estimates of the Higgs mass in heavy supersymmetry,''
  Phys.\ Rev.\ D {\bf 89} (2014) 5,  055023
  [arXiv:1312.5743 [hep-ph]];
  %%CITATION = ARXIV:1312.5743;%%
  %\bibitem{Kanemura:2014dja}
  S.~Kanemura, M.~Kikuchi and K.~Yagyu,
  %``Radiative corrections to the Yukawa coupling constants in two Higgs doublet models,''
  Phys.\ Lett.\ B {\bf 731} (2014) 27
  [arXiv:1401.0515 [hep-ph]];
  %%CITATION = ARXIV:1401.0515;%%
   %\bibitem{Kanemura:2015mxa}
  %S.~Kanemura, M.~Kikuchi and K.~Yagyu,
  %``Fingerprinting the extended Higgs sector using one-loop corrected Higgs boson couplings and future precision measurements,''
  Nucl.\ Phys.\ B {\bf 896} (2015) 80
  [arXiv:1502.07716 [hep-ph]];
  %%CITATION = ARXIV:1502.07716;%%
    S.~Kanemura, H.~Yokoya and Y.~J.~Zheng,
  %``Complementarity in direct searches for additional Higgs bosons at the LHC and the International Linear Collider,''
  Nucl.\ Phys.\ B {\bf 886} (2014) 524
  [arXiv:1404.5835 [hep-ph]];
  %%CITATION = ARXIV:1404.5835;%%    
%\bibitem{Anandakrishnan:2014qxa}
  A.~Anandakrishnan, B.~C.~Bryant and S.~Raby,
  %``Threshold Corrections to the Bottom Quark Mass Revisited,''
  JHEP {\bf 1505} (2015) 088
  [arXiv:1411.7035 [hep-ph]];
  %%CITATION = ARXIV:1411.7035;%%
  %\bibitem{Bae:2015nva}
  K.~J.~Bae, H.~Baer, N.~Nagata and H.~Serce,
  %``Prospects for Higgs coupling measurements in SUSY with radiatively-driven naturalness,''
  arXiv:1505.03541 [hep-ph].
  %%CITATION = ARXIV:1505.03541;%%

%%%%%%%%%%%%%%%%%%%%%%%%%%%%%%%
%%%%%%%%%%%%%%%%%%%%%%%%%%%%%%%

\bibitem{DHP}
%\bibitem{Dobado:1997up}
  A.~Dobado, M.~J.~Herrero and S.~Pe\~naranda,
  %``Decoupling of supersymmetric particles,''
  Eur.\ Phys.\ J.\ C {\bf 7} (1999) 313
  [hep-ph/9710313];
  %%CITATION = HEP-PH/9710313;%%
  %\bibitem{Dobado:1999cz}
  %A.~Dobado, M.~J.~Herrero and S.~Pe\~naranda,
  %``The SM as the quantum low-energy effective theory of the MSSM,''
  Eur.\ Phys.\ J.\ C {\bf 12} (2000) 673
  [hep-ph/9903211];
  %%CITATION = HEP-PH/9903211;%%
  %\bibitem{Dobado:2000pw}
  %A.~Dobado, M.~J.~Herrero and S.~Pe\~naranda,
  %``The Higgs sector of the MSSM in the decoupling limit,''
  Eur.\ Phys.\ J.\ C {\bf 17} (2000) 487
  [hep-ph/0002134].
  %%CITATION = HEP-PH/0002134;%%
  %58 citations counted in INSPIRE as of 29 May 2015
  
\bibitem{MSSM-Higgs_probes}
%\bibitem{Cahill-Rowley:2014wba}
  M.~Cahill-Rowley, J.~Hewett, A.~Ismail and T.~Rizzo,
  %``Higgs boson coupling measurements and direct searches as complementary probes of the phenomenological MSSM,''
  Phys.\ Rev.\ D {\bf 90} (2014) 9,  095017
  [arXiv:1407.7021 [hep-ph]];
  %%CITATION = ARXIV:1407.7021;%%
  %7 citations counted in INSPIRE as of 01 Jul 2015
%\bibitem{Carena:2014nza}
  M.~Carena  {\it et al.}, 
%H.~E.~Haber, I.~Low, N.~R.~Shah and C.~E.~M.~Wagner,
  %``Complementarity between nonstandard Higgs boson searches and precision Higgs boson measurements in the MSSM,''
  Phys.\ Rev.\ D {\bf 91} (2015) 3,  035003
  [arXiv:1410.4969 [hep-ph]];
  %%CITATION = ARXIV:1410.4969;%%
    %\bibitem{Bhattacherjee:2015sga}
  B.~Bhattacherjee, A.~Chakraborty and A.~Choudhury,
  %``Status of MSSM Higgs Sector using Global Analysis and Direct Search Bounds, and Future Prospects at the HL-LHC,''
  arXiv:1504.04308 [hep-ph].
  %%CITATION = ARXIV:1504.04308;%%
  
\bibitem{hMSSM}
%\bibitem{Djouadi:2013vqa}
  A.~Djouadi and J.~Quevillon,
  %``The MSSM Higgs sector at a high $M_{SUSY}$: reopening the low tan$\beta$ regime and heavy Higgs searches,''
  JHEP {\bf 1310} (2013) 028
  [arXiv:1304.1787 [hep-ph]];
  %%CITATION = ARXIV:1304.1787;%%
 %\bibitem{Djouadi:2013uqa}
  A.~Djouadi  {\it et al.}, 
%L.~Maiani, G.~Moreau, A.~Polosa, J.~Quevillon and V.~Riquer,
  %``The post-Higgs MSSM scenario: Habemus MSSM?,''
  Eur.\ Phys.\ J.\ C {\bf 73} (2013) 2650
  [arXiv:1307.5205 [hep-ph]];
  %%CITATION = ARXIV:1307.5205;%%
  %\bibitem{Djouadi:2015jea}
  A.~Djouadi  {\it et al.}, 
%L.~Maiani, A.~Polosa, J.~Quevillon and V.~Riquer,
  %``Fully covering the MSSM Higgs sector at the LHC,''
  arXiv:1502.05653 [hep-ph].
  %%CITATION = ARXIV:1502.05653;%%
  
%\cite{Khachatryan:2014jba}
\bibitem{Khachatryan:2014jba}
  V.~Khachatryan {\it et al.}  [CMS Collaboration],
  %``Precise determination of the mass of the Higgs boson and tests of compatibility of its couplings with the standard model predictions using proton collisions at 7 and 8 $\,\text {TeV}$,''
  Eur.\ Phys.\ J.\ C {\bf 75} (2015) 5,  212
  [arXiv:1412.8662 [hep-ex]];
  %%CITATION = ARXIV:1412.8662;%%
  %105 citations counted in INSPIRE as of 04 Jun 2015
%\bibitem{ATLAS-CONF-2015-007}
  The ATLAS collaboration,
  %``Measurements of the Higgs boson production and decay rates and coupling strengths using pp collision data at √s = 7 and 8 TeV in the ATLAS experiment,''
  ATLAS-CONF-2015-007, ATLAS-COM-CONF-2015-011.
  %%CITATION = ATLAS-CONF-2015-007, ATLAS-COM-CONF-2015-011;%%
 
\bibitem{accuracy-refs}  
%\bibitem{Dawson:2013bba}
  S.~Dawson {\it et al.}, 
%A.~Gritsan, H.~Logan, J.~Qian, C.~Tully, R.~Van Kooten, A.~Ajaib and A.~Anastassov {\it et al.},
  %``Working Group Report: Higgs Boson,''
  arXiv:1310.8361 [hep-ex];
  %%CITATION = ARXIV:1310.8361;%%
  %150 citations counted in INSPIRE as of 26 May 2015  
%\bibitem{Englert:2014uua}
  C.~Englert {\it et al.}, 
%A.~Freitas, M.~M.~M\"uhlleitner, T.~Plehn, M.~Rauch, M.~Spira and K.~Walz,
  %``Precision Measurements of Higgs Couplings: Implications for New Physics Scales,''
  J.\ Phys.\ G {\bf 41} (2014) 113001
  [arXiv:1403.7191 [hep-ph]];
  %%CITATION = ARXIV:1403.7191;%%
  %47 citations counted in INSPIRE as of 29 May 2015
%\cite{Moortgat-Picka:2015yla}
%\bibitem{Moortgat-Picka:2015yla}
  G.~Moortgat-Pick {\it et al.}, 
%H.~Baer, M.~Battaglia, G.~Belanger, K.~Fujii, J.~Kalinowski, S.~Heinemeyer and Y.~Kiyo,
  %``Physics at the $e^+ e^-$ Linear Collider,''
  arXiv:1504.01726 [hep-ph].
  %%CITATION = ARXIV:1504.01726;%%
  %3 citations counted in INSPIRE as of 26 May 2015

%\cite{Zeppenfeld:2000td}
\bibitem{ZeppenfeldGianotti}
  D.~Zeppenfeld, R.~Kinnunen, A.~Nikitenko and E.~Richter-Was,
  %``Measuring Higgs boson couplings at the CERN LHC,''
  Phys.\ Rev.\ D {\bf 62} (2000) 013009
  [hep-ph/0002036];
  %%CITATION = HEP-PH/0002036;%%
  %270 citations counted in INSPIRE as of 24 Jun 2015
%\bibitem{Gianotti}
  F.~Gianotti and M.~Pepe-Altarelli,
  %``Precision physics at the LHC,''
  Nucl.\ Phys.\ Proc.\ Suppl.\  {\bf 89} (2000) 177
  [hep-ex/0006016].
  %%CITATION = HEP-EX/0006016;%%
  %22 citations counted in INSPIRE as of 24 Jun 2015

%\cite{Barlow:2004wg}
\bibitem{Barlow:2004wg}
  R.~Barlow,
  %``Asymmetric statistical errors,''
  physics/0406120; 
  %%CITATION = PHYSICS/0406120;%%
  %``Asymmetric errors,''
  eConf C {\bf 030908} (2003) WEMT002 [physics/0401042 [physics.data-an]];
  %%CITATION = PHYSICS/0401042;%%
  %16 citations counted in INSPIRE as of 11 Jun 2015
  %\bibitem{Barlow:2003sg}
  %R.~Barlow,
  %``Asymmetric systematic errors,''
  physics/0306138;
  %%CITATION = PHYSICS/0306138;%%
  {\tt \url{http://www.slac.stanford.edu/~barlow/statistics.html}}.
  
%\cite{Heinemeyer:1998yj}
\bibitem{Heinemeyer:1998yj}
  S.~Heinemeyer, W.~Hollik and G.~Weiglein,
  %``FeynHiggs: A Program for the calculation of the masses of the neutral CP even Higgs bosons in the MSSM,''
  Comput.\ Phys.\ Commun.\  {\bf 124} (2000) 76
  [hep-ph/9812320].
  %%CITATION = HEP-PH/9812320;%%
  %888 citations counted in INSPIRE as of 29 May 2015

%\cite{Carena:2013qia}
\bibitem{Carena:2013qia}
  M.~Carena {\it et al.}, 
%S.~Heinemeyer, O.~Stal, C.~E.~M.~Wagner and G.~Weiglein,
  %``MSSM Higgs Boson Searches at the LHC: Benchmark Scenarios after the Discovery of a Higgs-like Particle,''
  Eur.\ Phys.\ J.\ C {\bf 73} (2013) 9,  2552
  [arXiv:1302.7033 [hep-ph]].
  %%CITATION = ARXIV:1302.7033;%%
  %109 citations counted in INSPIRE as of 18 Jun 2015
  
%\cite{Cahill-Rowley:2013gca}
\bibitem{Cahill-Rowley:2013gca} 
  M.~W.~Cahill-Rowley {\it et al.}, 
%J.~L.~Hewett, A.~Ismail, M.~E.~Peskin and T.~G.~Rizzo,
  %``pMSSM Benchmark Models for Snowmass 2013,''
  arXiv:1305.2419 [hep-ph].
  %%CITATION = ARXIV:1305.2419;%%

%\cite{Agashe:2014kda}
\bibitem{PDG}
  K.~A.~Olive {\it et al.}  [Particle Data Group Collaboration],
  %``Review of Particle Physics,''
  Chin.\ Phys.\ C {\bf 38} (2014) 090001.
  %%CITATION = CHPHD,C38,090001;%%
  %1155 citations counted in INSPIRE as of 29 May 2015

%\cite{Heinemeyer:2000nz}
\bibitem{Heinemeyer:2000nz}
  S.~Heinemeyer, W.~Hollik and G.~Weiglein,
  %``FeynHiggsFast: A Program for a fast calculation of masses and mixing angles in the Higgs sector of the MSSM,''
  hep-ph/0002213.
  %%CITATION = HEP-PH/0002213;%%
  %139 citations counted in INSPIRE as of 29 May 2015
  
%\cite{Altmannshofer:2012ks}
\bibitem{Altmannshofer:2012ks}
  W.~Altmannshofer, M.~Carena, N.~R.~Shah and F.~Yu,
  %``Indirect Probes of the MSSM after the Higgs Discovery,''
  JHEP {\bf 1301} (2013) 160
  [arXiv:1211.1976 [hep-ph]].
  %%CITATION = ARXIV:1211.1976;%%
  %67 citations counted in INSPIRE as of 04 Jun 2015

\bibitem{MAtanb_exclusion_refs}  
%\cite{Khachatryan:2014wca}
%\bibitem{Khachatryan:2014wca}
  V.~Khachatryan {\it et al.}  [CMS Collaboration],
  %``Search for neutral MSSM Higgs bosons decaying to a pair of tau leptons in pp collisions,''
  JHEP {\bf 1410} (2014) 160
  [arXiv:1408.3316 [hep-ex]];
  %%CITATION = ARXIV:1408.3316;%%
  %\bibitem{Aad:2014vgg}
  G.~Aad {\it et al.}  [ATLAS Collaboration],
  %``Search for neutral Higgs bosons of the minimal supersymmetric standard model in pp collisions at $\sqrt{s}$ = 8 TeV with the ATLAS detector,''
  JHEP {\bf 1411} (2014) 056
  [arXiv:1409.6064 [hep-ex]].
  %%CITATION = ARXIV:1409.6064;%%
  %32 citations counted in INSPIRE as of 26 May 2015

\end{thebibliography}

\end{document}